\documentclass{esannV2}
\usepackage[dvips]{graphicx}
\usepackage[utf8]{inputenc} 
\usepackage{amssymb,amsmath,array}

\newcommand{\R}{\mathbb{R}}
\usepackage{hyperref}
\usepackage{caption}
\usepackage{subcaption}
\usepackage{color}

\begin{document}
\title{A nuclear-norm based convex formulation\\for informed source separation}

\author{Augustin Lef\`evre$^1$, Fran\c cois Glineur$^{1,2}$ and P.-A. Absil$^1$
%
\thanks{This paper presents research results of the Belgian Network DYSCO
(Dynamical Systems, Control, and Optimization), funded by the
Interuniversity Attraction Poles Programme, initiated by the Belgian
State, Science Policy Office. The scientific responsibility rests with
its authors.}
%
\vspace{.3cm}\\
%
1- Universit\'e catholique de Louvain -  ICTEAM Institute\\
Avenue Georges Lema\^itre 4, B-1348 Louvain-la-Neuve - Belgium
%
\vspace{.1cm}\\
2- Universit\'e catholique de Louvain - CORE\\
Voie du Roman Pays 34, B-1348 Louvain-la-Neuve - Belgium\\
}

\maketitle

\begin{abstract}
We study the problem of separating audio sources from a single linear mixture. The goal is to find a decomposition of the single channel spectrogram into a sum of individual contributions associated to a certain number of sources. In this paper, we consider an informed source separation problem in which the input spectrogram is partly annotated. We propose a convex formulation that relies on a nuclear norm penalty to induce low rank for the contributions. We show experimentally that solving this model with a simple subgradient method outperforms a previously introduced nonnegative matrix factorization (NMF) technique, both in terms of source separation quality and computation time.
\end{abstract}

\section{Introduction}

Single-channel source separation is an underdetermined problem, commonly used as a pre-processing technique for higher-level tasks (speech recognition in complex environments, polyphonic music transcription, etc.). In this paper, we consider an annotated problem, where partial information on the sources is available. 

While exact source recovery cannot be expected in general, a key ingredient in source separation techniques consists in assuming some form of redundancy in the data, which renders the problem overdetermined. This is typically done by assuming that the source contributions have low rank, which leads to non-convex formulations. The nonnegative matrix factorization model was first applied to audio signals for polyphonic transcription\cite{SmBr03}, and was earlier introduced in other contexts\cite{Pata94}.

In this article, we investigate a novel convex formulation of the source separation problem, where low rank of the contributions is induced with nuclear norm penalty terms.

\section{Low-rank linear models for audio source separation}

Single-channel source separation consists in recovering a certain number of unknown source signals from linear measurements of their sum. In the case of audio separation, signals consist of spectrograms, i.e.\@ matrices of coefficients in the time-frequency domain, with the property that the spectrogram of the mixed signal is the sum of the spectrograms of the individual sources. These spectrograms are obtained by way of time-frequency transforms \cite{dsp_oppenheimer} to enhance redundancy in the data, see Figure \ref{stft} below : a Fourier transform is computed on short time segments of the signal. The phase is then discarded to yield approximate translation invariance. 

\begin{figure}[htdp]
\centering
\includegraphics[width=0.7\linewidth]{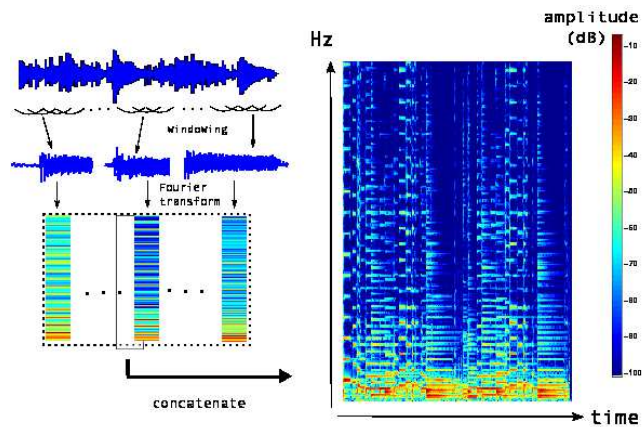}
\caption{Time-frequency operators enhance sparsity in audio signals}
\label{stft}
\end{figure}

Single-channel source separation methods can benefit strongly from prior knowledge on the sources. State-of-the-art methods consist in finding a factorization of the spectrogram into a product of matrices, one consisting in elementary spectra, called the dictionary, and the other representing activation coefficients of those spectra. Columns of the dictionary matrix are then grouped manually by an expert into sources. 

\begin{figure}[h!]
\centering\includegraphics[width=0.4\linewidth]{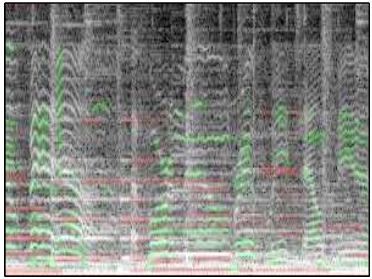}
\caption{Annotations in the time-frequency plane to improve source separation}
\label{annot}
\end{figure}

In this paper, we consider a different situation in which some coefficients of the source spectrograms are known to be equal to some pre-specified targets. For example, Figure \ref{annot} depicts annotations in a two-source scenario, such as one where the input audio signal mixes a singer's voice with an accompanying instrument. A user has annotated $20 \%$ of the time-frequency plane, i.e.\@ has identified some
time-frequency regions, shown in red or green, where one of the two sources dominates the other. This allows to specify target values for the source estimates of each signal for these annotated coefficients.

\section{Formulations for informed source separation}
\subsection{Informed source separation}
We formalize the annotated audio source separation problem as follows. Our input matrix is $\bar{V} \in \R_+^{F \times N}$, where $F$ denotes the number of frequency bins, and $N$ the number of time bins. 
We assume that  $\bar{V} \approx \sum_{g=1}^G V^{(g)} $, where the $G$ sources to be identified $V^{(g)} \in \R_+^{F \times N}$ are assumed to be low-rank matrices. 

The set of annotated observations is denoted by $\mathcal{L}$, a subset of $\{1,\dots, F\}\times \{1,\dots,N\}$. For each time-frequency bin $(f,n) \in \mathcal{L}$, a target value $\hat{V}^{(g)}_{f n}$ is provided for each source $g$, according to \[ \hat{V}^{(g)}_{f n}=M^{(g)}_{f n} \bar{V}_{f n} \quad \forall 1 \le g \le G \quad \forall(f,n) \in \mathcal{L} \] where masking coefficients $M_{f n}^{(g)}$ satisfy equality $\sum_{g=1}^G M_{f n}^{(g)}=1$  $\forall(f,n) \in \mathcal{L}$. 

\subsection{A formulation based on nonnegative matrix factorization}
The informed source separation was considered in \cite{LEBAFE12}, where it is solved by way of a modified nonnegative matrix factorization (NMF) problem. More specifically, the power spectrogram of source $g$ is modelled as $V^{(g)}=D^{(g)} A^{(g)}$, i.e.\@ the low-rank product of nonnegative factors $D^{(g)} \in \R_+^{F \times K}$ with the corresponding nonnegative activation coefficients $A^{(g)} \in \R_+^{K \times N}$. Note that the rank $K$ for each source spectrogram needs to be fixed in advance. The following formulation 
\begin{equation*}
\begin{array}{cc}
\min & D_{IS}(\bar{V},\sum_{g=1}^G D^{(g)} A^{(g)}) +\lambda \sum_{(f,n) \in \mathcal{L}} \sum_{g=1}^G D_{IS}(\hat{V}^{(g)}_{f n},[D^{(g)} A^{(g)}]_{f n}) \\ \text{s.t.} &
D^{(g)} \geq 0 \text{ and } A^{(g)} \geq 0 \,. 
\end{array}
\end{equation*}
is then solved with multiplicative updates, a standard algorithm for NMF. The first term of the objective function measures dissimilarity between input spectrogram and its reconstruction with the Itakura-Saito divergence $D_{IS}$. The second term penalizes deviations for the annotated coefficients (as these constraints cannot be enforced exactly with NMF algorithms). The resulting problem is nonconvex and multimodal, so that only local minima can be computed. In practice, one obtains good source estimates by starting from many initial points and selecting the best solutions, at the cost of increasing the computing time.

\subsection{A convex formulation based on a nuclear norm penalty}
We propose a novel convex formulation whose variables are the source spectrograms $V^{(g)}$, which allows us to impose the annotation constraints $ V^{(g)}_{f n} = \hat{V}^{(g)}_{f n}$ exactly. The rank of those spectrograms is no longer fixed in advance, but is instead minimized by way of a nuclear-norm based penalty. Denoting the nuclear norm by $\|X\|_\star$, i.e.\@ the sum of singular values of $X$, our formulation is
\begin{equation}
\begin{array}{cc}
\min & \|\bar{V} - \sum_{g=1}^G V^{(g)} \|_F^2 + \lambda \sum_{g=1}^G \|V^{(g)}\|_\star \,\\
\text{subject to} & V^{(g)}\ge 0 \text{ and } V^{(g)}_{f n} = \hat{V}_{f n} \ \forall (f,n) \in \mathcal{L}\quad 1\leq g\leq G,\end{array}
\label{lownuc}
\end{equation}

Dissimilarity between the input spectrogram and its approximation is now measured with a Frobenius norm, chosen mostly for convenience. Note that source spectrograms are only required to have nonnegative coefficients, a condition weaker than the nonnegative factorization used in the previous formulation. Our experiments (see Section~\ref{sec:Exp}) show that this has no consequence on later post-processing steps and allows to capture more complex models of source signals at no additional cost. 

This model is convex, hence in principle easier to solve than the NMF formulation, with algorithms computing solutions that are globally optimal for the problem. It is however also nonsmooth, because of the nuclear norm penalties. It is well-known that nuclear-norm based formulations can be recast as semidefinite programs (SDP), see \cite{ReFaPa10}. However, interior-point solvers applicable to SDP are limited to problems with a relatively small size (less than a hundred of frequency and time bins), which is insufficient for our audio application. We use instead a first-order method applied directly to our formulation, namely a classical projected subgradient scheme. The simple structure of the constraints (nonnegativity and fixed values) ensures that projections are easy to compute, and a subgradient of the objective function can be computed relatively cheaply. Indeed, its first term is smooth (with a simple gradient), and a subgradient can be obtained for each nuclear norm term at the cost of computing a singular value decomposition.


Choosing the step size in subgradient schemes is not a trivial task. For simplicity, we choose a decreasing step size rule of the form $\alpha_t=\frac{\alpha_0}{1+t}$, whose convergence to the minimum is guaranteed~\cite[Th.~2.3]{Sho1985}. Nevertheless, obtaining high accuracy solutions can be often very slow, and we stop the algorithm after a fixed number of iterations.

In the next section, we compare our new approach with NMF, focusing mostly on the quality of the source estimates, in order to validate our new convex formulation. The design and implementation of a more efficient nonsmooth convex optimization method is left for further research.

\section{Numerical experiments}\label{sec:Exp}
\subsection{Experimental setup}

We compare our approach (referred to as ``lownuc'') with the NMF formulation of \cite{LEBAFE12}, in controlled experimental conditions where the true sources are known. We can therefore measure the quality of the source estimates for each formulation\footnote{Track by track results, as well as listening tests, will be made available online at \url{www.di.ens.fr/~lefevrea/lownuc.html}.}. We fix the proportion of annotations to $40 \%$. 
Once source spectrograms have been estimated, the corresponding audio signal is computed and its quality is assessed by computing Signal-to-Distortion, Signal-to-Interference, and Signal-to-Artefact ratios. These quantities, expressed in dB, vary from $-\infty$ (for a null estimate) to $+\infty$ (for a perfect estimate). 
As source signals in our experiments have equal $\ell_2$ norm in each audio track, so that providing the mixed signal as a guess for any source yields $0$ SDR.

Both approaches feature some hyperparameters: $\lambda$ for the penalty strength in both formulations, rank $K$ of the source spectrograms for NMF and initial step length $\alpha_0$ for the subgradient technique. In order to present a fair comparison, we tried several representative values for each hyperparameter and selected for each test problem and each formulation the model with the best SDR\footnote{Selecting hyperparameters in unsupervised settings is still an open problem in statistical learning.}.  Both methods were run with the same CPU time budget of 180 seconds.

\subsection{Tests on the SISEC database}
\begin{figure}
\centering
\begin{subfigure}{.40 \linewidth}
		\includegraphics[width=\linewidth]{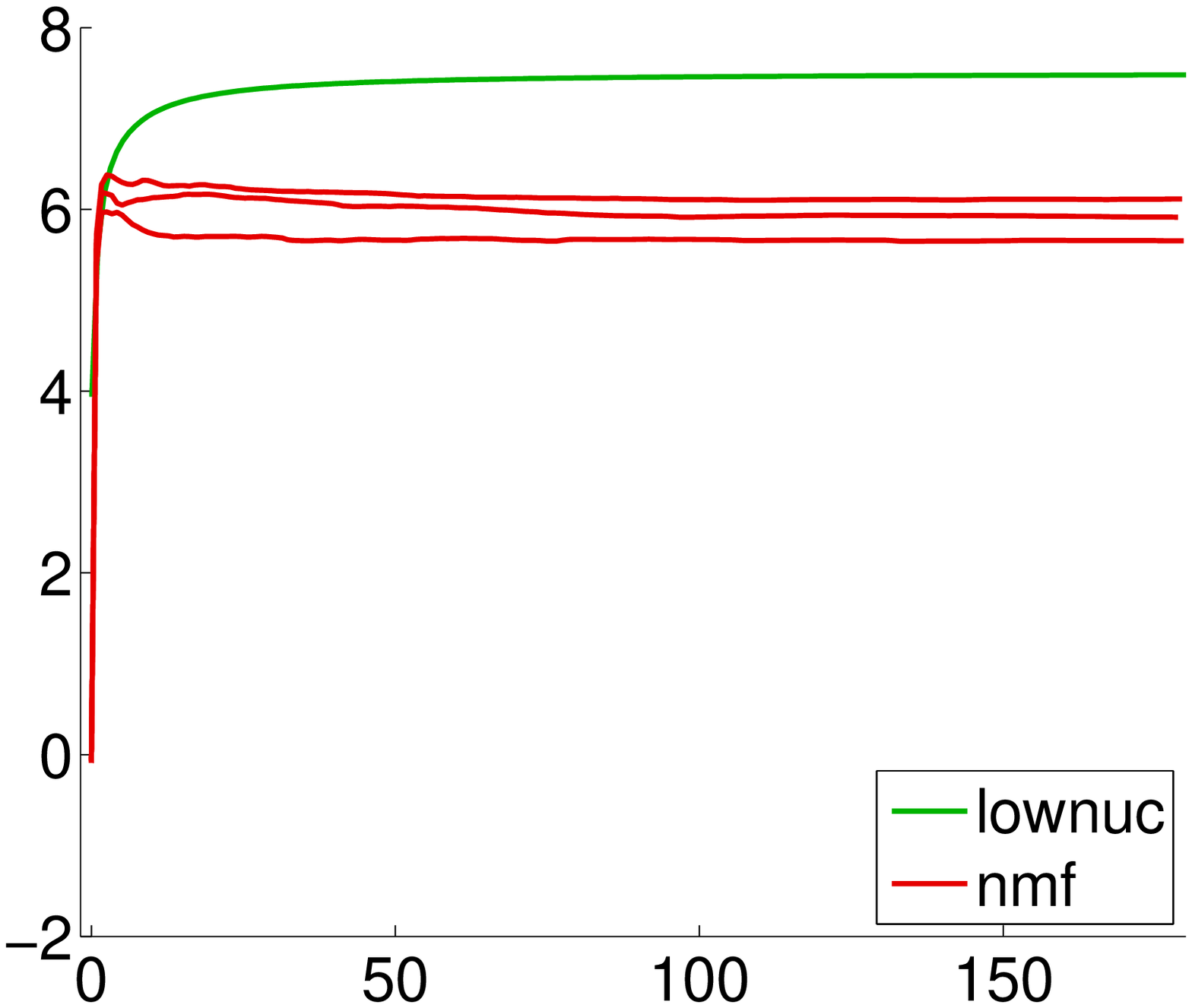}
		\end{subfigure}
\begin{subfigure}{.4 \linewidth}
		\includegraphics[width=\linewidth]{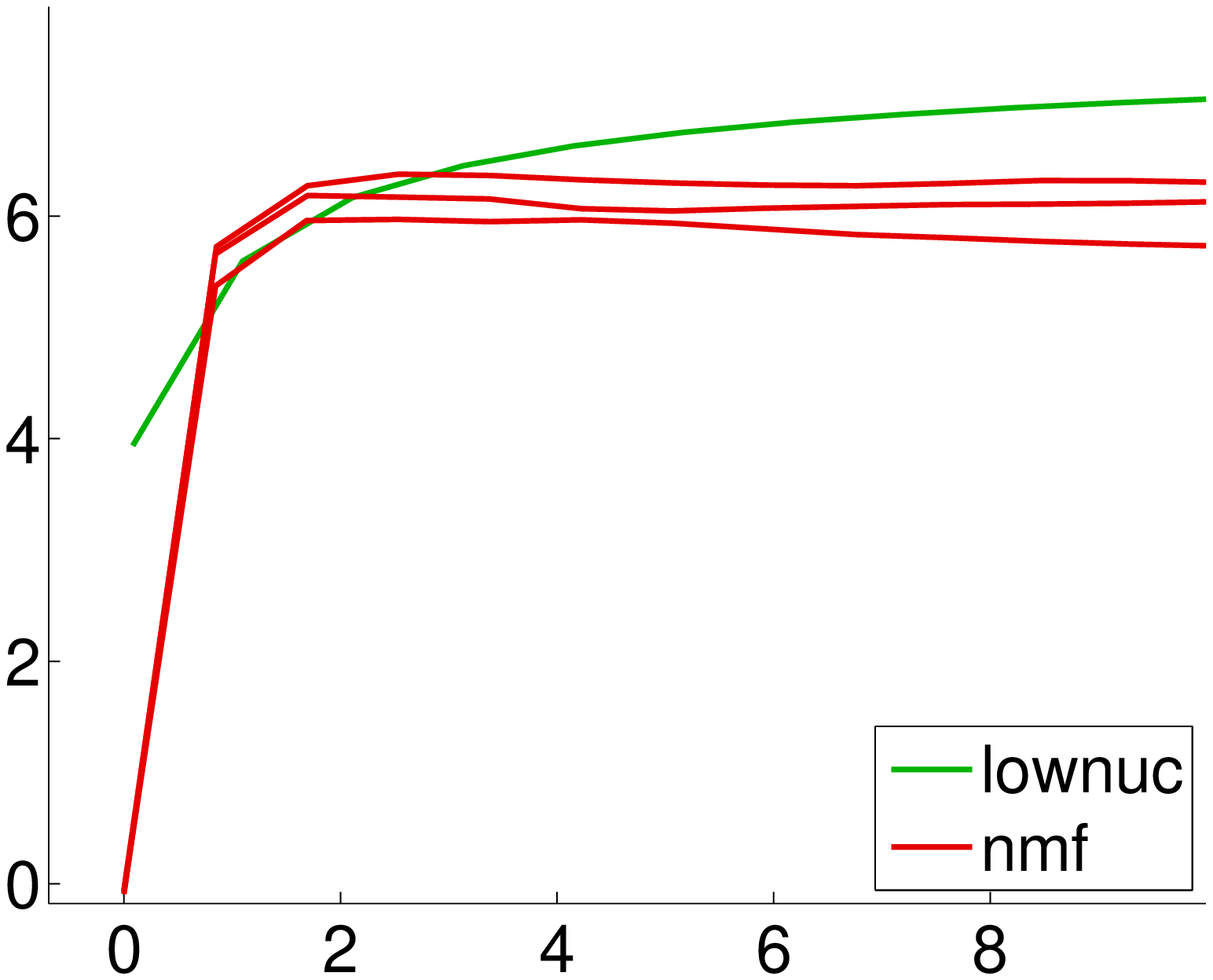}
		\end{subfigure}		
		\caption{(Left) Evolution of SDR as a function of CPU time (in seconds), for (\textcolor{blue}{blue}) our method and \textcolor{red}{(red)} NMF started from several initial points. (Right) Zoom on the first few seconds.}		\label{sdrtime}
\end{figure}

The SISEC database (Professionally produced music recordings track) consists in 5 tracks, each 14 seconds' long. All tracks were downsampled to 16 kHz. $512$ samples-long analysis windows were used to compute spectrograms, with $256$ samples of overlap. Each spectrogram has roughly $10^6$ entries. Each track is composed of two sources, voice and accompaniment.
We include in our comparison the  ``oracle'' estimates (computed using the true values of the source spectrograms, representing the best possible accuracy) as well as the so-called ``lazy'' estimates (projections of the uninformative estimates $V^{(g)}=\frac{1}{G} \bar{V}$ on the constraints of Problem~\eqref{lownuc}). 
As we can see in Table \ref{global_results}, both formulations improve substantially over lazy estimates, with our approach beating NMF by roughly $0.85$ dB on average SDR. Despite the simplicity of the subgradient scheme, our approach is also attractive in terms of computing time, as illustrated on Figure \ref{sdrtime}(a). A closer look at the first few seconds of each run (Figure \ref{sdrtime}(b)) shows that it improves over NMF as soon as the CPU time budget allows for more than ten seconds of computations.

\begin{table}[htdp]
\centering
{\footnotesize
\begin{tabular}{|l|l|l|l|}
\hline
&\textbf{SDR}&\textbf{SIR}&\textbf{SAR}\\\hline
lazy&3.4725&4.9059&10.2163\\\hline
nmf&7.9267&16.1891&8.8206\\\hline
lownuc&8.779&16.0186&9.9494\\\hline
oracle&10.8523&19.1113&11.6088\\\hline
\end{tabular}

}
\caption{Average results on SISEC database using $40 \%$ of annotations.}
\label{global_results}
\end{table}

\vspace*{-.5cm}
\section{Conclusion and perspectives} 

We have introduced a convex formulation of informed source separation using low-rank inducing penalties. Preliminary results show that our approach performs favorably in comparison with a previous formulation based on NMF, both in terms of source separation quality and computing time. Besides improving the efficiency of the resolution of the new formulation (e.g.\@ using a smoothing technique), an interesting direction for future research would consist in using non-Euclidean dissimilarity measure, whose use is known to be crucial as the amount of annotations decreases. In particular, convex dissimilarity measures such as those used in \cite{plca_07} would fit naturally into our framework. Finally, robustness to wrong or inaccurate annotations is another worthy goal to pursue.


\begin{footnotesize}
\bibliographystyle{unsrt}
\bibliography{bib_esann,bib_esann_PA}

\begin{thebibliography}{1}

\bibitem{SmBr03}
P.~Smaragdis and J.C. Brown.
\newblock Non-negative matrix factorization for polyphonic music transcription.
\newblock In {\em IEEE Workshop on Applications of Signal Processing to Audio
  and Acoustics (WASPAA)}, 2003.

\bibitem{Pata94}
P.~Paatero and U.~Tapper.
\newblock Positive matrix factorization: A non-negative factor model with
  optimal utilization of error estimates of data values.
\newblock {\em Environmetrics}, 1994.

\bibitem{dsp_oppenheimer}
A.~Oppenheimer and R.~Schafer.
\newblock {\em Digital signal processing}.
\newblock Prentice-Hall, 1975.

\bibitem{LEBAFE12}
A.~Lef{\`e}vre, F.~Bach, and C.~F{\'e}votte.
\newblock Semi-supervised {NMF} with time-frequency annotations for
  single-channel source separation.
\newblock In {\em International Conference on Music Information Retrieval
  (ISMIR)}, 2012.

\bibitem{ReFaPa10}
B.~Recht, M.~Fazel, and P.A. Parrilo.
\newblock {Guaranteed Minimum-Rank Solutions of Linear Matrix Equations via
  Nuclear Norm Minimization}.
\newblock {\em SIAM Review.}, 2010.

\bibitem{Sho1985}
N.~Z. Shor.
\newblock {\em Minimization methods for nondifferentiable functions}, volume~3
  of {\em Springer Series in Computational Mathematics}.
\newblock Springer-Verlag, Berlin, 1985.
\newblock Translated from the Russian by K. C. Kiwiel and A. Ruszczy{\'n}ski.

\bibitem{plca_07}
P.~Smaragdis, B.~Raj, and M.V. Shashanka.
\newblock Supervised and semi-supervised separation of sounds from
  single-channel mixtures.
\newblock In {\em International Conference in Independant Component Analysis
  (ICA)}, 2007.

\end{thebibliography}
\end{footnotesize}

\end{document}